\pdfoutput=1
\documentclass[aps,prc,superscriptaddress,twocolumn]{revtex4-2}

\usepackage{graphicx}
\usepackage{color}
\usepackage{amsmath,amssymb,bm,mathtools}
\usepackage{braket}
\usepackage{physics}
\usepackage{hyperref}
\hypersetup{hidelinks}
\usepackage{enumitem}
\usepackage{booktabs}
\usepackage{tensor}
\usepackage{mathrsfs}
\usepackage{xspace}
\usepackage{siunitx}

\newcommand{\rvec}{\bm{r}}

\newcommand{\up}{{\!\uparrow}}
\newcommand{\down}{{\!\downarrow}}

\newcommand{\fm}{{\mathrm{fm}}}

\newcommand{\vari}[2]{\frac{\delta {#1}}{\delta {#2}}}

\begin{document}
\date{\today}
\title{Neural-Network-Based Variational Method in Nuclear Density Functional Theory: Application to the Kohn--Sham method}

\author{Kenta Yoshimura}
\affiliation{Department of Physics, School of Science, Institute of Science Tokyo, Tokyo 152-8550, Japan}

\author{Kazuyuki Sekizawa}
\email{sekizawa@phys.sci.isct.ac.jp}
\affiliation{Department of Physics, School of Science, Institute of Science Tokyo, Tokyo 152-8550, Japan}
\affiliation{Nuclear Physics Division, Center for Computational Sciences, University of Tsukuba, Ibaraki 305-8577, Japan}
\affiliation{RIKEN Nishina Center, Saitama 351-0198, Japan}

\begin{abstract}
We extend the neural-network-based variational method for nuclear density functional theory to the Kohn--Sham scheme, representing the complex spinor components of the single-particle orbitals by multi-layer perceptrons. 
We show that neural-network optimization of a given energy density functional, combined with an orthonormalization post-processing step, is mathematically equivalent to the variational condition projected onto the tangent space of the wave-function manifold spanned by the network parameters, and that the training optimizes not only the expansion coefficients but also the basis functions themselves. 
We assess the method from three points of view. 
In the first place, we examine how the results depend on the number of units, the number of layers, and the arithmetic precision, and find that quantitative accuracy requires both a sufficient width and a sufficient depth, while single-precision arithmetic is sufficient to represent the nuclear density distribution. 
In the second place, the binding energies and charge radii of several closed-shell nuclei agree with conventional Skyrme--Hartree--Fock results, and the quadrupole deformations of open-shell nuclei are consistent with reference calculations that include pairing and with experiment. 
In the third place, we confirm that a neural-network single-particle basis can represent the three-dimensional configurations of the fundamental pasta phases: spheres, rods, and slabs.
The framework offers a new perspective on computational nuclear theory, well suited to the forthcoming generation of GPU- and AI-oriented high-throughput supercomputers.

\end{abstract}
\maketitle
\section{Introduction}
\label{sec:introduction}

Nuclear density functional theory (DFT) is one of the most powerful theoretical tools for describing a variety of many-body phenomena across widely different length scales, from the structures and dynamics of individual nuclei to the properties of neutron star matter~\cite{hohenberg1964dft, kohn1965ks, drut2010dftreview, nakatsukasa2016,Bender2003,Colo2020,Yang2020}.
The conceptual foundation of DFT is the Hohenberg--Kohn (HK) theorem, which guarantees the existence of an energy density functional (EDF) whose minimization yields the ground state of the many-body system.
The Kohn--Sham (KS) scheme then enables a Hartree--Fock-like formulation of DFT by introducing auxiliary single-particle orbitals in order to incorporate shell effects as well as deformations.
Additionally, to fully describe higher-order deformation modes~\cite{scamps2021deform3d}, various fission and fusion dynamics~\cite{umar2006fusion3d, goddard2015fission3d, bulgac2016fission3d}, cluster configurations~\cite{ichikawa2011cluster3d, ebran2014cluster3d}, and nuclear pasta phases in neutron stars~\cite{newton2009pasta3dft, schuetrumpf2013pasta3d, schuetrumpf2019}, it is essential to perform three-dimensional mesh calculations~\cite{maruhn2014sky3d, pei2014hfb3d, jin2021lise, chen2022hfb3d}, which are generally computationally expensive and require supercomputer resources.

At the same time, the computational environment has been changing drastically in recent years.
Modern supercomputers are designed around modern GPU and AI-accelerator platforms, allowing high-throughput tensor operations~\cite{markidis2018tensorcore}, automatic differentiation~\cite{baydin2018autodiff}, and the use of low- or mixed-precision arithmetic~\cite{higham2022mixedprecision, Tan2023}.
Although existing nuclear DFT implementations have achieved high numerical accuracy and several codes already exploit GPU acceleration~ \cite{jin2021lise}, they are commonly based on double-precision arithmetic.
Such implementations remain executable on current accelerator-based systems, but may not fully exploit the substantially higher throughput of forthcoming AI-oriented supercomputers.
We can therefore expect the DFT framework to be expressible in a form that couples more directly to the machine-learning architecture, by minimizing the EDF itself as the loss function without the iterative self-consistent diagonalization required in conventional Kohn--Sham solvers.
Motivated by this, we proposed a ``neural-network-based variational method''~\cite{yoshimura2026nnetf} for the extended Thomas--Fermi (ETF) or orbital-free formulation of nuclear DFT, where the neutron and proton densities at each spatial point are represented by multi-layer perceptrons (MLPs).
In that work, we showed that the proposed framework successfully describes the structures and properties of several finite nuclei, as well as some fundamental nuclear pasta phases.

In the present work, we extend this neural-network (NN) approach to the Kohn--Sham method.
Whereas the orbital-free ETF formulation cannot capture the shell structure of nuclei, the Kohn--Sham scheme restores it through the introdution of single-particle orbitals, at the cost of a more involved numerical treatment.
From these orbitals, we can construct the number, kinetic, and other densities that constitute the EDF.
This structure suggests that the neural network can equally well represent the single-particle orbitals themselves.
Indeed, neural-network wave functions have been applied successfully to many-body Slater determinants in both electronic~\cite{Carleo2017, Pfau2020, Hermann2020, li2023d4ft, tan2024nndft} and nuclear~\cite{Adams2021, Lovato2022, fore2025nnpasta, wen2026neuralquantumstates, wu2026mesonnucleusboundstatesneuralnetwork} systems.
Although in those studies the ground states were obtained by Monte Carlo methods, the underlying variational property is shared with the Kohn--Sham method.
However, replacing the KS orbitals by the NN output raises several numerical issues.
First, the trial functions must satisfy the orthonormality condition.
Moreover, realistic Skyrme calculations with spin-orbit coupling require the treatment of complex spinors and the associated kinetic and spin-current densities.
These requirements make the present Kohn--Sham application a suitable test for validating our proposed NN-based method.

In this study, we demonstrate how the NN-based variational method connects to the theoretical basis of conventional nuclear DFT, and we assess its numerical validity from three viewpoints.
In the first place, we examine how the results depend on the numbers of units, layers and on the arithmetic precision.
In the second place, we compare the binding energies and charge radii of several closed-shell nuclei with those obtained from conventional calculations, and examine whether deformation emerges spontaneously in open-shell nuclei.
In the third place, we apply the method to nuclear pasta phases and verify that the three fundamental shapes, namely sphere, rod, and slab, can be represented by the NN ansatz.
Through these test cases, we clarify both the theoretical foundation and the numerical performance of the NN-based approach.

The article is organized as follows.
Section~\ref{sec:formalism} presents the theoretical basis of the NN-based method and the Skyrme EDF employed in this work.
In Sec.~\ref{sec:results}, the results of the NN-based variational calculations for finite nuclei and nuclear pasta are presented and the feasibility of the present formalism is discussed.
In Sec.~\ref{sec:summary}, we summarize this article and discuss future prospects.

\section{Formalism}
\label{sec:formalism}

\subsection{Neural-orbital variational problem}
\label{sec:nn_orbitals}

We consider the space spanned by the single-particle orbitals of neutrons and protons.
The spin degree of freedom is treated explicitly as a two-component spinor $\psi_{q,i}(\bm{r}) = \qty(\psi_{q,i}(\bm{r}\,\up), \psi_{q,i}(\bm{r}\,\down))^\text{T}$.
The occupied orbitals form the following Slater determinant:
\begin{equation}
    \ket{\Phi}
    =
    \prod_{q=n,p}
    \prod_{i=1}^{N_q}
    a_{q,i}^\dagger \ket{0},
    \label{eq:slater_determinant}
\end{equation}
The single-particle wave functions are subjected to the orthonormality condition,
\begin{equation}
    \sum_{\sigma}
    \int \dd\rvec\,
    \psi_{q,i}^*(\rvec\sigma)\psi_{q,j}(\rvec\sigma)
    =
    \delta_{ij},
    \label{eq:orbital_orthonormality}
\end{equation}
where $\sigma$ labels the spin component.

For a Skyrme-type EDF, the total energy of the system is written as a functional of various local densities, such as the particle number density $\rho$ and the kinetic density $\tau$, all of which are constructed from the single-particle orbitals. The EDF can be written formally as
\begin{equation}
    E = E[\rho,\tau, \ldots] = E[\{\psi_{q,i}\}].
    \label{eq:orbital_functional}
\end{equation}
Taking the variation of Eq.~\eqref{eq:orbital_functional} under the constraint on the orthonormality condition~\eqref{eq:orbital_orthonormality}, we obtain
\begin{equation}
    \frac{\delta E}{\delta \psi_{q,i}^*(\rvec)}
    =
    \epsilon^{(q)}_i\psi_{q,i}(\rvec),
    \label{eq:orbital_variation}
\end{equation}
where $\epsilon^{(q)}_i$ is a Lagrange multiplier arising from the constraint.

We now examine how this is expressed under the NN ansatz.
Suppose that each single-particle wave function $\tilde{\psi_{q}}$ is represented by an MLP as
\begin{equation}
    \tilde{\psi}_{q,i}(\bm{r}) = g_\text{MLP}(\bm{r},\theta_q).
\end{equation}
Here the tilde indicates that no processing such as orthonormalization has yet been applied; through suitable post-processing we obtain a set of single-particle orbitals $\psi_{q,i}$ that satisfy the orthonormality condition~\eqref{eq:orbital_orthonormality}.
Identifying the loss function with the energy (plus correction terms), we have
\begin{equation}
    \min_{\theta_p,\theta_n}
    \mathcal L(\theta_p,\theta_n),
    \qquad
    \mathcal L = E[\{\psi_{q,i}(\theta_q)\}] + \mathcal L_{\mathrm{ext}},
    \label{eq:nn_energy_minimization}
\end{equation}
where $\mathcal{L}_\text{ext}$ is an appropriate constraint term chosen according to the system.
In this way, the variational problem of DFT can be recast as an optimization problem in machine learning.
This optimization condition is expressed as the requirement that the gradient of the energy with respect to $\theta_q$ vanishes:
\begin{equation}
    \pdv{E}{\theta_{q,k}}
    =
    \sum_i
    \int \dd\rvec\,
    \frac{\delta E}{\delta\psi_{q,i}^*(\rvec)}
    \pdv{\psi_{q,i}^*(\rvec)}{\theta_{q,k}}
    + \mathrm{c.c.}
    =
    0.
    \label{eq:projected_orbital_stationarity}
\end{equation}
Furthermore, assuming that the orbitals $\psi_{q,i}$ always satisfy the orthonormality condition, so that
\begin{equation}
    \int \dd\bm{r} \, \vari{E}{\psi^*_{q,i}(\bm{r})} = 0
\end{equation}
holds, it follows that
\begin{equation}
    \sum_i
    \int \dd\rvec\,
    \qty(\frac{\delta E}{\delta\psi_{q,i}^*(\rvec)} - \epsilon^{(q)}_i\psi_{q,i}(\rvec))
    \pdv{\psi_{q,i}^*(\rvec)}{\theta_{q,k}}
    + \mathrm{c.c.}
    =
    0.
    \label{eq:tangent_space_projection}
\end{equation}
That is, the energy-optimization problem under the NN ansatz is equivalent to projecting the variational condition~\eqref{eq:orbital_variation} onto the tangent space of the manifold spanned by the wave functions that the parameters represent.
If the NN has sufficient expressive power, this solution should coincide with the exact solution in the full function space.
Consequently, the mathematical validity of our NN ansatz is ensured.

Let us now look more closely at the internal structure of the NN.
The NN used in this work takes a spatial coordinate as input and returns the values of all the orbitals at that point.
By construction, the number of output orbitals is determined by the particle number of the system.
We denote this number by $N_\text{orbit}$ and the number of units in the hidden layers by $N_\text{unit}$.
The architecture of the NN is then
\begin{equation}
    3 \to N_\text{unit}\to \cdots \to N_\text{orbit}.
\end{equation}
The expressive power of the NN also depends on the number of hidden layers, which we denote by $L$ hereafter.

We collect the unit values of each layer into a vector $\bm{h}^{(\ell)}$.
The operation associated with each hidden layer is then written as
\begin{align}
    \bm{h}^{(0)}(\bm{r}) &= \bm{r},\\
    \bm{h}^{(\ell)}(\bm{r})&= f(W^{(\ell)}_q\bm{h}^{(\ell-1)}(\bm{r}) + \bm{b}^{(\ell)}_q),\quad \ell = 1,\ldots ,L,
\end{align}
where $f$ is the activation function, $W^{(\ell)}_q$ is the weight matrix, and $\bm{b}^{(\ell)}_q$ is the bias vector.
The transformation from the last hidden layer to the output layer is parametrized by $W_q^\text{out}$ and $\bm b_q^\text{out}$.
While the outputs of an NN are usually real, the wave functions are in general complex; the output of a given orbital therefore has four components, namely, the real and imaginary parts of the two spinor components.
Labeling these components by $\mu\in\{\Re\,\up,\ \Im\,\up,\ \Re\,\down,\ \Im\,\down\}$, we write the network output as
\begin{equation}
    \tilde\psi_{q,i}^{(\mu)}(\rvec)
    =
    \sum_{k=1}^{N_{\mathrm{unit}}}
        (W_q^{\mathrm{out}})_{i\mu,k}\, h_k^{(L)}(\rvec)
    + (b_q^{\mathrm{out}})_{i\mu}.
    \label{eq:spinor_output}
\end{equation}
The parameter set $\theta_q$ comprises the weights $W_q^{(\ell)}$ and $W_q^\text{out}$, together with the biases $\bm b_q^{(\ell)}$ and $\bm b_q^\text{out}$, of all layers.

Next, we examine how the orbitals are optimized through the training of the parameters.
We write the parameters of the hidden layers as $\theta^\text{hid}_q=\{ W^{(\ell)}_q, \bm{b}^{(\ell)}_q\}$.
Since the output of the last hidden layer $\bm{h}^{(L)}(\bm{r})$ is built from these parameters, we may write
\begin{equation}
    h_k^{(L)}(\bm{r}) = \chi_k (\bm{r};\theta^\text{hid}_q),
\end{equation}
where $k$ is the unit index.
The output layer, which represents the single-particle orbitals, is then written formally as
\begin{equation}
    \tilde\psi_{q,i}(\rvec)
    =
    \sum_{k=1}^{N_{\mathrm{unit}}}
        \bm{c}_{q,ik}\,\chi_k(\rvec;\theta_q^{\mathrm{hid}}),
    \label{eq:adaptive_basis}
\end{equation}
where the bias term has been absorbed into the coefficient matrix $\bm{c}_{q,ik}$.
If the hidden-layer parameters are fixed at given values $\hat{\theta}^{\text{hid}}_q$, the optimization in the NN reduces simply to an optimization of the coefficients over the space spanned by the fixed basis $\{\chi_k(\rvec;\hat{\theta}_q^{\mathrm{hid}})\}$.
It must be noted, however, that this basis is not necessarily orthonormal.
The training of the hidden-layer parameters, on the other hand, changes the basis \textit{itself}.
That is, the training of the NN wave functions can be separated into two processes: optimization of the basis functions themselves, and optimization of the coefficients that expand the orbitals in that basis.
In other words, there are two ways to improve the computational accuracy: (i) increasing the number of units, thereby increasing the number of basis functions, and (ii) increasing the number of hidden layers, thereby enhancing the expressive power of the basis.
As one would intuitively expect, these two effects are closely intertwined, which will be demonstrated by the results presented in Sec.~\ref{sec:results}.

\subsection{Skyrme energy density functional}
\label{sec:skyrme_edf}

As introduced in Sec.~\ref{sec:nn_orbitals}, the Skyrme EDF is a functional of the particle number, kinetic, and spin-current densities,
\begin{equation}
    E_{\mathrm{tot}} = E[\rho, \tau, J] ,
    \label{eq:energy_decomposition}
\end{equation}
where we impose the time-reversal symmetry (TRS) and drop the time-odd terms, which is valid static calculations of even-even nuclei.
These densities are evaluated from the orthonormal single-particle orbitals as follows:
\begin{align}
    \rho_q(\rvec)
    &=
    \sum_{i,\sigma}
    |\psi_{q,i}(\rvec\sigma)|^2 ,
    \label{eq:rho_spinor} \\
    \tau_q(\rvec)
    &=
    \sum_{i,\sigma}
    |\grad \psi_{q,i}(\rvec\sigma)|^2 ,
    \label{eq:tau_spinor} \\
    J_{q,\mu\nu}(\rvec)
    &=
    \operatorname{Im}\qty[
    \sum_{i,ss'}
    \psi_{q,i}^{*}(\rvec s)
    (\sigma_\mu)_{ss'}
    \nabla_\nu\psi_{q,i}(\rvec s')] .
    \label{eq:J_spinor}
\end{align}
Here $\bm{\sigma}=(\sigma_x,\sigma_y,\sigma_z)$ denotes the Pauli matrices in the spin space, while $s$ and $s'$ label the spin indices.
For any density $X\in \{\rho,\tau, J\}$, the isoscalar and isovector components are defined as
\begin{equation}
    X_0 = X_n + X_p,\quad X_1 = X_n - X_p .
\end{equation}
The spin-orbit part of the EDF involves the antisymmetric vector component of the spin-current density, given by
\begin{equation}
    J^\lambda_q(\bm{r}) = \sum_{\mu\nu} \epsilon_{\lambda\mu\nu} J_{q,\mu\nu}(\bm{r}) .
\end{equation}
These densities are defined at each point in coordinate space, and the energy is expressed in terms of them.

The total energy of a general system can be decomposed into the kinetic term, the interaction term, and the Coulomb term as
\begin{equation}
    E_\text{tot} = E_\text{kin}^\text{eff} + E_\text{int} + E_\text{Coul} .
\end{equation}
The effective kinetic energy $E^\text{eff}_\text{kin}$ is obtained from the bare kinetic energy
\begin{equation}
    E_\text{kin} = \sum_q \int \dd\bm{r}\, \frac{\hbar^2}{2m_q}\, \tau_q(\bm{r})
\end{equation}
by subtracting the center-of-mass correction $E_\text{CM}$, that is,
\begin{equation}
    E_\text{kin}^\text{eff} = E_\text{kin} - E_\text{CM} .
\end{equation}
In this work we adopt the one-body approximation for the center-of-mass correction,
\begin{equation}
    E_\text{CM} \approx \frac{E_\text{kin}}{A} ,
\end{equation}
where $A= N_n + N_p$ is the mass number.
The interaction term is written, for a Skyrme EDF, as
\begin{equation}
    \begin{aligned}
        E_\text{int} &= \int \dd\bm{r}\, \sum_{t=0,1}\Big[C^\rho_t [\rho_0]\rho_t^2 + C^{\Delta\rho}_t\rho_t\Delta\rho_t \\&\hspace{12mm} + C^\tau_t \rho_t\tau_t + C^{\nabla J}_t\rho_t \grad \cdot\bm{J}_t\Big] .
    \end{aligned}
\end{equation}
This retains only the time-even components of the central part and the spin-orbit (LS) part.
Here, $C^{X}_t$ (with $X=\{\rho,\Delta\rho,\tau,\nabla J\}$ and isospin index $t=\{0,1\}$) collectively denotes the coupling constants of the Skyrme EDF; for their detailed definitions we refer the reader to, \textit{e.g.}, Ref.~\cite{lesinski2007skyrme}.

The Coulomb term is written as the sum of a direct term and an exchange term:
\begin{equation}
    E_{\mathrm{Coul}}
    =
    \frac{1}{2}\int \dd\rvec\,\rho_p(\rvec)V_\text{C}(\rvec)
    -
    \frac{3}{4}
    \left(\frac{3}{\pi}\right)^{1/3}
    e^2
    \int \dd\rvec\,\rho_p^{4/3}(\rvec),
    \label{eq:coulomb_edf}
\end{equation}
where the exchange term is evaluated in the Slater approximation.
The Coulomb potential $V_\text{C}(\bm{r})$ contained in the direct term is obtained as the solution of the Poisson equation,
\begin{equation}
    \grad^2 V_\text{C}(\bm{r}) = -4\pi e^2 \rho_\text{ch}(\bm{r}) .
\end{equation}
This equation is solved with a fast Fourier transform (FFT) algorithm.
In this work, we neglect the charge form factor and approximate the charge density as $\rho_\text{ch} = \rho_p$.
For an isolated nucleus, to avoid the spurious contributions introduced by periodicity, the FFT is performed on an enlarged grid and a truncated kernel is used to suppress the periodic images~\cite{jin2021lise}.
In neutron-star matter, electrons uniformly fill the cell so as to satisfy the charge-neutrality condition $\int \dd\bm{r}\,(\rho_p(\bm{r}) - \rho_e)=0$; with $V$ the volume of the cell, the electron density is $\rho_e = N_p/V$ and the charge density is $\rho_\text{ch} = \rho_p - \rho_e$.
In this case the DC term vanishes automatically, so that no truncated kernel is needed.
For a system containing electrons, the electron contribution $E_e$ should in principle be added to the total energy of the system; however, for a fixed input $N_p$ this term enters merely as a constant and does not affect the variation.
For this reason, we neglect the electron energy in the present work to simplify the calculation.

\subsection{Computational settings}
\label{sec:numerics}

In all calculations in the present study, spatial functions are represented on a cell-centered cubic Cartesian grid $[-x_{\max},x_{\max}]^3$ with $N$ points per axis, spacing $\dd x=2x_{\max}/N$, and volume element $\dd V=\dd x^3$; no spatial symmetry is imposed, so that the solver retains full triaxial freedom.
Unless stated otherwise, we use $N=48$ grid points per axis and choose $x_{\max} = 12$~fm.
For neutron-star matter, one should in principle search for the optimal cell size that minimizes the energy; here, for simplicity, we fix the cell size for each system.
Spatial derivatives are evaluated using a spectral method with a FFT algorithm.

For the MLP architecture, we typically use $N_\text{unit}=128$ units, a hidden-layer depth of $L=4$, and $\tanh$ as the activation function.
For finite nuclei, in order to concentrate the distribution at the center of the cell, we multiply the orbitals by an envelope function of sigmoid form,
\begin{equation}
    \begin{aligned}
        g_{\mathrm{env}}(r)
        &=
        \operatorname{sigmoid}
        \left(
            -\frac{r-r_{\mathrm{cut}}}{a_{\mathrm{cut}}}
        \right),\\
        \tilde\psi_{q,i}(\rvec)
        &\rightarrow
        g_{\mathrm{env}}(|\rvec|)\tilde\psi_{q,i}(\rvec),
    \end{aligned}
    \label{eq:finite_envelope}
\end{equation}
thereby suppressing the tails that would otherwise leak to the cell boundary.
This procedure serves only to stabilize the calculation numerically; we have checked that, once the cutoff radius $r_\text{cut}$ is taken sufficiently large, the results do not depend on $r_\text{cut}$.
For the neutron-star matter calculations, on the other hand, no such treatment is applied, since the density distribution of dripped neutrons must be described.

\begin{table*}[t]
    \centering
    \caption{Dependence of the $^{16}$O results on the network architecture and on the
    arithmetic precision, obtained with SLy4 including the spin-orbit terms.
    Listed are the energy per nucleon, the charge radius, the proton quadrupole moment, and the
    average wall time per variational step.
    The fp32 and fp64 halves are obtained under identical mesh and optimization settings.}
    \label{tab:stage1_sweep}
    \begin{tabular}{cc@{\hspace{1.5em}}
        S[table-format=-1.4]S[table-format=1.3]S[table-format=1.3]S[table-format=2.0]
        @{\hspace{1.5em}}
        S[table-format=-1.4]S[table-format=1.3]S[table-format=1.3]S[table-format=2.0]}
        \toprule
        & & \multicolumn{4}{c}{fp32} & \multicolumn{4}{c}{fp64} \\
        \cmidrule(lr){3-6}\cmidrule(lr){7-10}
        $L$ & $N_{\text{unit}}$ &
        {$E/A$ [MeV]} & {$R_{\text{ch}}$ [fm]} & {$|Q_p|$ [fm$^2$]} & {Time [ms/step]} &
        {$E/A$ [MeV]} & {$R_{\text{ch}}$ [fm]} & {$|Q_p|$ [fm$^2$]} & {Time [ms/step]} \\
        \midrule
        2 & 16  & -7.9005 & 2.809 & 0.259 & 20 & -7.8997 & 2.809 & 0.267 & 18 \\
        2 & 32  & -7.8392 & 2.852 & 0.583 & 17 & -7.8381 & 2.853 & 0.634 & 16 \\
        2 & 64  & -7.9802 & 2.796 & 0.028 & 18 & -7.9801 & 2.796 & 0.027 & 18 \\
        2 & 128 & -8.0192 & 2.786 & 0.008 & 20 & -8.0192 & 2.786 & 0.008 & 18 \\
        2 & 256 & -8.0256 & 2.784 & 0.007 & 20 & -8.0255 & 2.784 & 0.007 & 22 \\
        \midrule
        4 & 16  & -7.9761 & 2.792 & 0.096 & 19 & -7.9754 & 2.793 & 0.099 & 18 \\
        4 & 32  & -8.0225 & 2.784 & 0.066 & 21 & -8.0226 & 2.784 & 0.064 & 20 \\
        4 & 64  & -8.0294 & 2.782 & 0.014 & 22 & -8.0293 & 2.782 & 0.014 & 20 \\
        4 & 128 & -8.0305 & 2.782 & 0.001 & 19 & -8.0304 & 2.782 & 0.001 & 22 \\
        4 & 256 & -8.0307 & 2.782 & 0.000 & 26 & -8.0307 & 2.782 & 0.000 & 28 \\
        \bottomrule
    \end{tabular}
\end{table*}

The raw wave-function values $\tilde{\psi}_{q,i}$ obtained from the output layer are, in general, not orthonormal, so a post-processing step is required.
This orthonormalization is performed by a Cholesky factorization of the overlap matrix (see Appendix~\ref{sec:otherformula} for details).
However, we note that the orbitals obtained after this orthonormalization,
\begin{equation}
    \tilde{\psi}_{q,i}(\bm{r}) \to \psi_{q,i}(\bm{r}) ,
\end{equation}
merely span an orthonormal basis and are not necessarily eigenstates of the mean-field potential, but rather they are linear combinations of such eigenstates.
Nevertheless, using the optimized orbitals together with their respective densities, we can transform the NN orbitals into the canonical (eigen)states.
This point is discussed in Appendix~\ref{sec:canonical_orbitals}.

The optimization process consists mainly of the following two phases.
The first is a pretraining step, in which the distribution is guided into a shape suited to the computational setup to provide a reasonable starting point for the subsequent minimization.
For finite nuclei, we use a Woods--Saxon density distribution,
\begin{equation}
    f_{\mathrm{WS}}(\rvec)
    =\Big[1+\exp\!\big((\abs{\bm{r}}-R_0)/a_{\mathrm{WS}}\big)\Big]^{-1},
    \label{eq:ws_target}
\end{equation}
with $R_0=1.2A^{1/3}$~fm and $a_{\mathrm{WS}}=0.67$~fm, normalized to the proton and neutron numbers $Z$ and $N$.
For deformed nuclei, in order to break the rotational symmetry, we impose a nuclide-dependent quadrupole deformation $\beta_2$ by setting $R_0 \to R_0\qty(1 + \beta_2 Y_{20})$.
In this phase, instead of the energy of the system, the loss function is taken to be the sum of squared deviations from the above target density, and this pretraining is run for 1000 steps.
For the pasta phases, instead of pretraining, the target shape is induced by adding to the loss function a guiding-potential constraint,
\begin{equation}
    \mathcal{L}_\text{ext} = \int \dd\bm{r}\, V_\text{guide}(\bm{r})\rho_0(\bm{r}),
\end{equation}
suited to the targeted structure, which is applied for 30000 steps.
Once this preparatory stage is complete, the optimization is carried out with the Skyrme EDF as the loss function, without imposing any further constraint.
For both the preparatory stage and the main training, we use Adam as the parameter optimizer; in the main training, after 4000 Adam steps, the calculation is refined with L-BFGS for more precise convergence, using a history size of 50.
The calculation is terminated once the energy change $\Delta E^{(n)} = E^{(n)} - E^{(n-1)}$ stays below $10^{-6}$~MeV for 150 consecutive steps.
Using the density distribution obtained after this calculation, we compute the total energy, charge radius, deformation, and other observables of the system.
Detailed expressions for the charge radius and other quantities are given in Appendix~\ref{sec:otherformula}.

All of the computational code is written in Python, and the training part is implemented using the PyTorch library.
Throughout every stage of the implementation, extension, and optimization of the code, we use Claude.ai by Anthropic, verifying the validity of the results at each step.
The calculations are performed on the supercomputer Miyabi, equipped with NVIDIA GH200, installed at the Information Technology Center of the University of Tokyo.

\section{Results}
\label{sec:results}

\subsection{Architecture and precision dependence}
\label{sec:architecture_results}

We first examine how the results vary with the number of hidden layers $L$, the number of units $N_\text{unit}$, and the arithmetic precision.
The calculations are performed for $^{16}$O with the SLy4 parametrization of the Skyrme EDF, including the spin-orbit terms.
Because spin is not a good quantum number once the spin-orbit interaction is included, each occupied state is represented as a full two-component spinor, so that $N_\text{orbit}=8$ for each nucleon species.
A $48^3$ spatial grid is used throughout.
After pretraining, the optimization is run for a fixed number of $30\,000$ steps, without applying the convergence criterion of Sec.~\ref{sec:numerics}; we then record the energy per nucleon $E/A$, the charge radius $R_\text{ch}$, the proton quadrupole moment $|Q_p|$, and the average wall time per step.
The results are collected in Table~\ref{tab:stage1_sweep}.
In the following, the deepest and widest calculation ($L=4$, $N_\text{unit}=256$) is taken as the reference against which the remaining entries are assessed.

\begin{table*}[t]
    \centering
    \caption{Energy per nucleon $E/A$ and charge radius $R_{\text{ch}}$ calculated with the
    proposed neural-network-based method, compared with the reference values of the previous
    study~\cite{chabanat1998sly}. The SLy4 and SkM$^*$ parametrizations are used.}
    \label{tab:spherical}
    \begin{tabular*}{0.8\linewidth}{@{\extracolsep{\fill}}ccccccccc}
        \toprule
        & \multicolumn{2}{c}{$E/A$ (SLy4) [MeV]}
        & \multicolumn{2}{c}{$E/A$ (SkM$^*$) [MeV]}
        & \multicolumn{2}{c}{$R_{\text{ch}}$ (SLy4) [fm]}
        & \multicolumn{2}{c}{$R_{\text{ch}}$ (SkM$^*$) [fm]} \\
        \cmidrule(lr){2-3}
        \cmidrule(lr){4-5}
        \cmidrule(lr){6-7}
        \cmidrule(lr){8-9}
        Nucleus
        & This work & Ref.
        & This work & Ref.
        & This work & Ref.
        & This work & Ref. \\
        \midrule
        $^{16}$O    & $-8.027$ & $-8.032$ & $-7.971$ & $-7.983$ & 2.783 & 2.779 & 2.788 & 2.787 \\
        $^{40}$Ca   & $-8.602$ & $-8.606$ & $-8.522$ & $-8.526$ & 3.497 & 3.493 & 3.503 & 3.499 \\
        $^{48}$Ca   & $-8.638$ & $-8.706$ & $-8.689$ & $-8.750$ & 3.514 & 3.511 & 3.506 & 3.504 \\
        $^{132}$Sn  & $-8.326$ & $-8.359$ & $-8.358$ & $-8.413$ & 4.724 & 4.715 & 4.717 & 4.705 \\
        $^{208}$Pb  & $-7.828$ & $-7.863$ & $-7.817$ & $-7.866$ & 5.505 & 5.498 & 5.501 & 5.492 \\
        \bottomrule
    \end{tabular*}
\end{table*}

First, the shallow network with $L=2$ (upper part of the table) behaves erratically when the number of units is small.
At $N_\text{unit}=16$ and $32$ the solution is underbound by $130$ and $190$~keV per nucleon, respectively, with respect to the reference value and acquires a spurious quadrupole moment of $|Q_p|=0.26$ and $0.58$~fm$^2$, although the ground state of $^{16}$O is spherical.
These two points are moreover not ordered monotonically, which indicates that the optimization itself becomes unstable, rather than that the accuracy is simply limited by the expressive power of the ansatz.
Beyond $N_\text{unit}=64$ the energy decreases monotonically, but even at $N_\text{unit}=256$ it remains $5$~keV per nucleon above the reference value, with no clear indication of convergence.

Second, for the deeper network with $L=4$ (lower part of the table), the results improve rapidly with $N_\text{unit}$ and are essentially converged by $N_\text{unit}=128$: the energy differs from the reference value by only $0.2$~keV per nucleon, the charge radius agrees to three decimal places, and the residual quadrupole moment is $|Q_p|\le 0.001$~fm$^2$.
At every value of $N_\text{unit}$, the $L=4$ network yields a lower energy than the $L=2$ one, and, apart from the anomalous $N_\text{unit}=32$ point discussed above, this difference shrinks as the network is widened.
These trends are consistent with the picture given in Sec.~\ref{sec:nn_orbitals}: the number of units sets how many basis functions are available, whereas the number of hidden layers controls the flexibility of the basis functions themselves.
Table~\ref{tab:stage1_sweep} shows that neither alone is sufficient for nuclear structure calculations, and that a moderate depth combined with a moderate width is the more economical choice.

Third, we turn to the arithmetic precision, for which the table compares runs using fp32 with those using fp64.
We note first that the Cholesky factorization used for the orthonormalization fails in fp32 owing to the accumulation of rounding errors, and is therefore carried out in fp64 (see Appendix~\ref{sec:otherformula} for details); the runs labeled fp32 are thus not performed entirely in single precision.
With this caveat, we find no significant difference between the two, either in the values of $E/A$ and $R_\text{ch}$ or in their trends with respect to $N_\text{unit}$ and $L$: the two precisions differ by at most about $1$~keV per nucleon, which is well below the residual error of the ansatz itself.
Single precision is therefore sufficient to represent the nuclear density distribution, provided that enough units and layers are used.

The wall time likewise shows no systematic dependence on the precision: the fp32 and fp64 timings differ by at most a few milliseconds in either direction, that is, within the scatter of the measurement.
Both increase only mildly with the network size, from about $18$~ms per step at $N_\text{unit}=16$ to about $27$~ms at $N_\text{unit}=256$.
At this problem size the step time is dominated by kernel-launch and memory-transfer overheads rather than by floating-point throughput, so reducing the width of the arithmetic brings no benefit.
For heavier nuclei, for neutron-star matter, or on substantially larger grids, where the arithmetic itself becomes the bottleneck, a difference between the two precisions is expected to appear.
This is also where the present formulation should benefit from forthcoming AI-oriented processors, on which low- and mixed-precision arithmetic is the natural operating point: keeping fp32 as the basic working precision leaves the method well positioned for such hardware.

\subsection{Finite nuclei}
\label{sec:finite_nuclei}

Next, we present the ground states of various nuclei from $^{16}$O to $^{238}$U computed with the NN-based method.
Since the present formulation does not include time-odd terms, we restrict ourselves to even--even nuclei, and we employ the two widely used functionals, SLy4~\cite{chabanat1998sly} and SkM$^*$~\cite{bartel1982skm}.
For all calculations, we use a $48^3$ grid, $L=4$ layers, $N_\text{unit}=128$ units, and fp64 arithmetic.
We verify the results from two viewpoints.
First, we check the reproducibility of the binding energy, charge radius, and related quantities for spherical, pairing-free closed-shell nuclei, for which spherically symmetric results are already available in Ref.~\cite{chabanat1998sly} and which we aim to reproduce.
Second, we compute several nuclei whose ground states are known to be deformed, in order to confirm that the NN-based method correctly describes the deformation.

\subsubsection{Spherical nuclei}

Table~\ref{tab:spherical} shows the results for spherical nuclei.
For the two functionals SLy4 and SkM$^*$, we compute the total energy and charge radius and compare them with the spherical calculations of Ref.~\cite{chabanat1998sly}.
The charge radii agree with the reference values to within $0.3\%$ for all nuclei and both functionals.
The energies per nucleon likewise agree to within $0.8\%$, but the agreement is not uniform: for $^{16}$O and $^{40}$Ca the deviation is below $0.15\%$, whereas it grows to $0.4$--$0.8\%$ for $^{48}$Ca, $^{132}$Sn, and $^{208}$Pb.
The deviation is moreover systematically one-signed, the present calculation being underbound in every case.
We attribute this trend to the finite size of the computational box: the heavier and more neutron-rich systems have more extended surfaces, so that a cubic cell of $x_{\max}=12$~fm truncates the outermost orbitals more severely.
Consistently with this interpretation, the deviation for $^{48}$Ca ($0.78\%$ with SLy4) is an order of magnitude larger than that for $^{40}$Ca ($0.05\%$), even though the two differ only by the addition of eight neutrons.
For every nucleus, the proton quadrupole moment $|Q_p|$ remains negligibly small, confirming that spherical solutions are obtained.
These results show that the proposed method correctly represents the nuclear density distribution, and that the residual accuracy is limited by the box size rather than by the NN ansatz itself.

\subsubsection{Deformed nuclei}

\begin{table}[t]
    \centering
    \caption{Quadrupole deformation $\beta_2$ of deformed even--even nuclei from the present
    neural-network-based solver with SLy4 and SkM$^*$, compared with reference calculations
    and with experiment. The reference calculations use SLy6 rather than SLy4 and include
    pairing (see text). Values in parentheses for $^{238}$U are read from the figures of
    Refs.~\cite{nesterenko2016uran, tolokonnikov2015uran} and are therefore approximate.}
    \label{tab:deformed}
    \begin{tabular*}{\linewidth}{@{\extracolsep{\fill}}cccccc}
        \toprule
        & \multicolumn{2}{c}{This work} & \multicolumn{2}{c}{Ref.} & \\
        \cmidrule(lr){2-3}\cmidrule(lr){4-5}
        Nucleus & SLy4 & SkM$^*$ & SLy6 & SkM$^*$ & Expt. \\
        \midrule
        $^{20}$Ne  & 0.55 & 0.53 & 0.56   & 0.37   & 0.72 \\
        $^{24}$Mg  & 0.53 & 0.52 & 0.54   & 0.49   & 0.61 \\
        $^{238}$U  & 0.27 & 0.28 & (0.29) & (0.29) & 0.29 \\
        \bottomrule
    \end{tabular*}
\end{table}

\begin{figure}[t]
    \centering
    \includegraphics[width=0.4\textwidth]{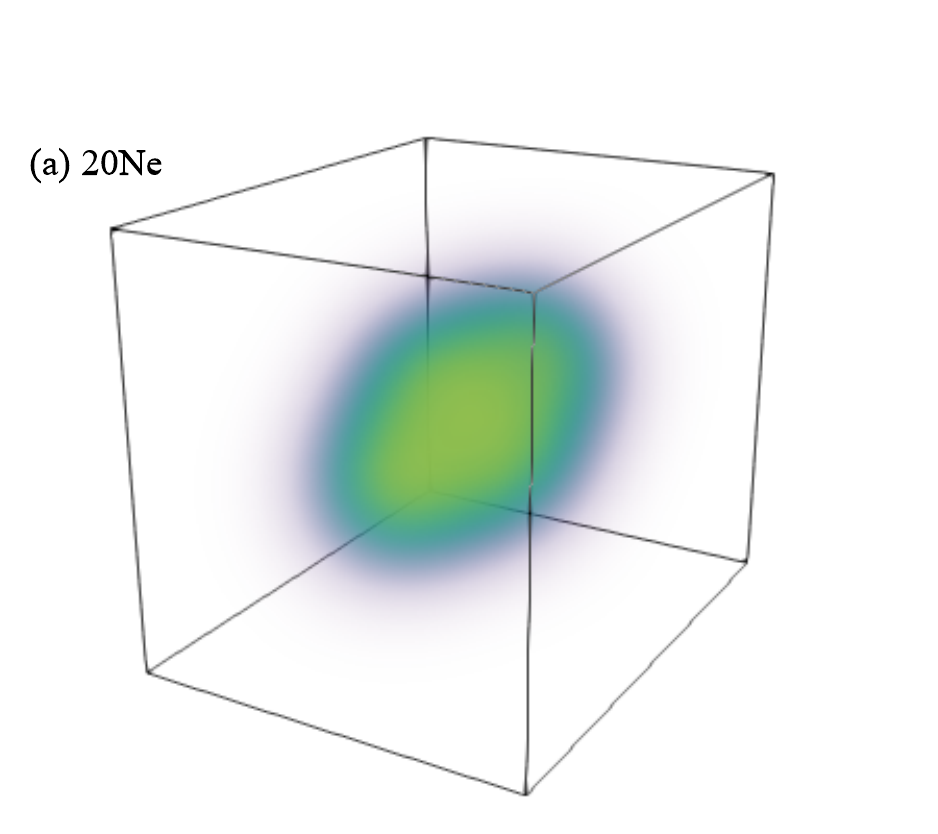}\\[-3.3mm]
    \includegraphics[width=0.4\textwidth]{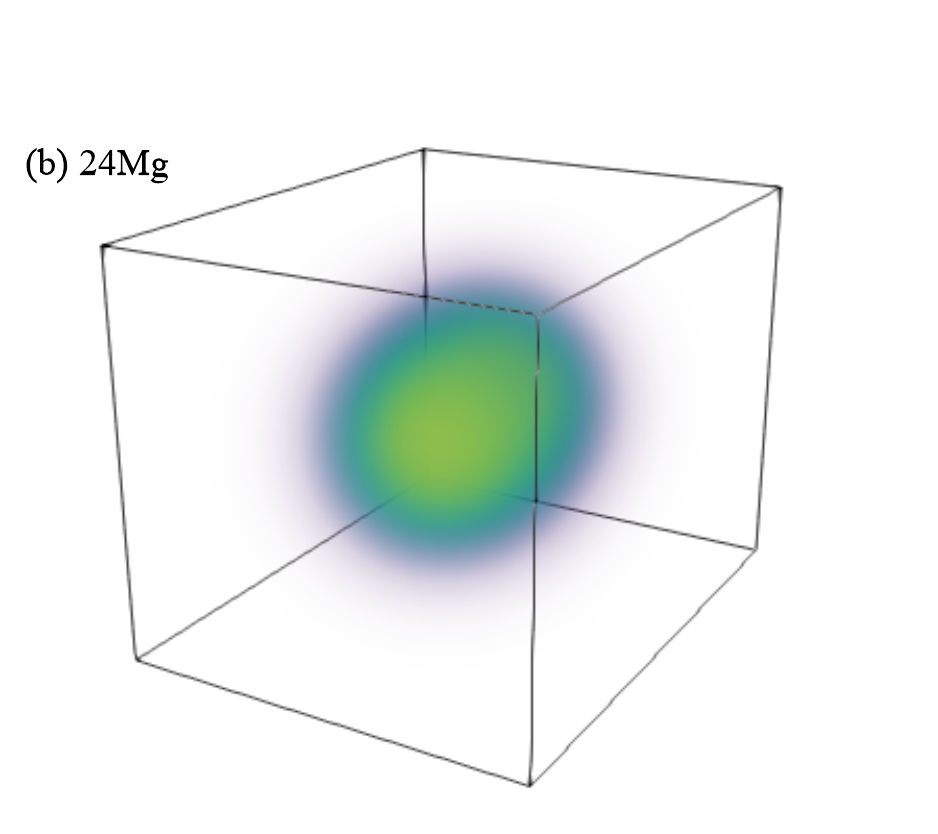}\\[-3.8mm]
    \includegraphics[width=0.4\textwidth]{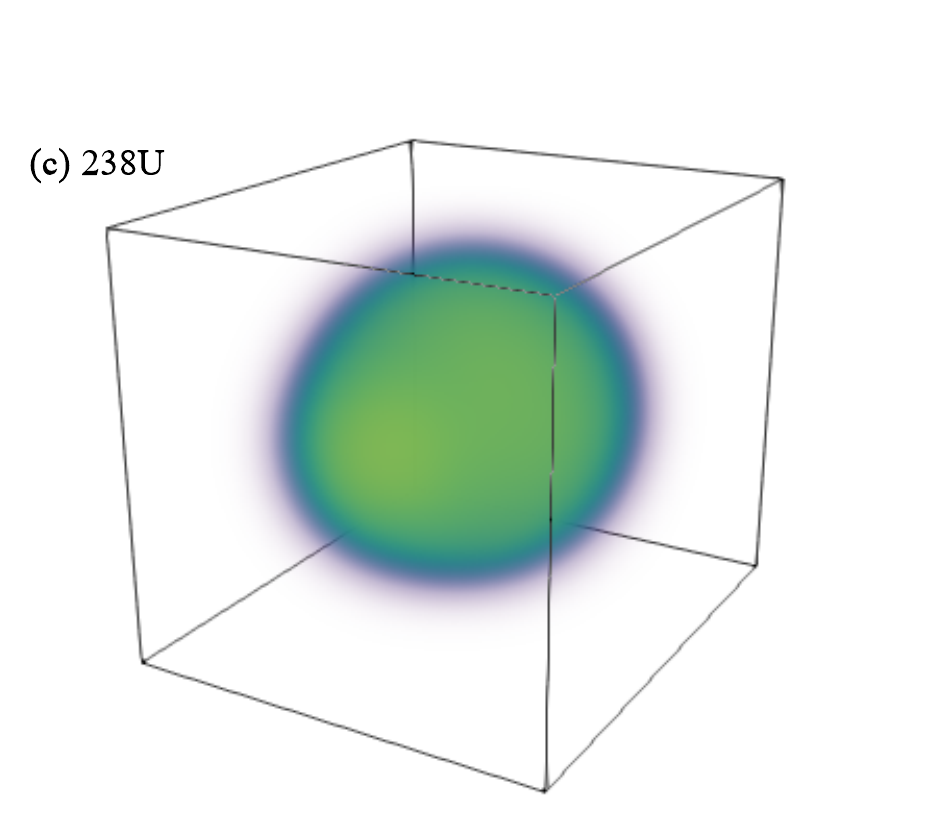}\vspace{-2mm}
    \caption{Nuclear density distributions of the deformed nuclei (a) $^{20}$Ne,
    (b) $^{24}$Mg, and (c) $^{238}$U, obtained with the neural-network method and the SkM$^*$
    functional.}
    \label{fig:deformed_density}
\end{figure}

Table~\ref{tab:deformed} presents calculations for several open-shell deformed nuclei, comparing the computed quadrupole deformations with reference values.
The experimental values are those deduced from the measured $B(E2)$ transition strengths~\cite{Raman2001}; the reference calculations for $^{20}$Ne and $^{24}$Mg are taken from Ref.~\cite{nesterenko2024ne-mg}, and those for $^{238}$U from Refs.~\cite{nesterenko2016uran, tolokonnikov2015uran}.
Two caveats apply to this comparison.
The reference results include a pairing interaction and were obtained with codes such as SkyAx~\cite{reinhard2021skyax} and HFBTHO~\cite{stoitsov2013hfbtho}, whereas the present calculations contain no pairing; and the reference SLy-family results were obtained with SLy6 rather than SLy4.
The comparison is therefore qualitative rather than quantitative.

With this in mind, the present calculations give $\beta_2=0.52$--$0.55$ for $^{20}$Ne and $^{24}$Mg and $\beta_2\approx0.27$--$0.28$ for $^{238}$U.
The SLy4 values agree closely with the SLy6 reference calculations for all three nuclei, the largest difference being $0.02$.
Agreement with the SkM$^*$ reference is good for $^{24}$Mg ($0.52$ against $0.49$) and for $^{238}$U, but for $^{20}$Ne the present value of $0.53$ substantially exceeds the reference value of $0.37$.
Since $^{20}$Ne is the lightest and least collective of the three, its deformation is the most sensitive to the pairing correlations that the present calculations omit, and we regard this difference as a plausible consequence of that omission rather than of the NN ansatz.
The experimental $\beta_2$ values are larger still for both $^{20}$Ne and $^{24}$Mg, as is expected for quantities extracted from $B(E2)$ strengths, which include dynamical contributions absent from a static mean-field calculation.
Overall, the method reproduces the deformed three-dimensional ground states at the level of accuracy that the comparison permits.

In Fig.~\ref{fig:deformed_density}, we show three-dimensional plots of the calculated density distributions of these three nuclei obtained with the SkM$^*$ functional.
For $^{238}$U in particular, a small but nonzero higher-order deformation is observed: the octupole deformation extracted from the density distribution is $\beta_3\approx0.03$.
Although this is smaller than the value $\beta_3=0.08$--$0.10$ inferred by recent studies~\cite{star2025octu, zhang2026octu}, it is noteworthy that such a deformation emerges spontaneously in an unconstrained three-dimensional NN-based calculation, arising purely from the energy optimization without imposing any shape (multipole) constraint.

\subsection{Nuclear pasta phases}

\label{sec:pasta}
Finally, we apply the NN-based method to the nuclear pasta phases inside neutron stars.
In the inner crust of neutron stars, it has long been known that the competition between the Coulomb energy and the surface energy gives rise to nuclear structures of various shapes, such as rods and slabs~\cite{Ravenhall1983, Hashimoto1984}, and calculations based on three-dimensional meshes have been carried out~\cite{schuetrumpf2019, nakamura2026ofdft}.
In our previous study~\cite{yoshimura2026nnetf}, we showed that densities represented by a neural network describe these structures well.
Whether the same holds when the network represents single-particle orbitals provides a stringent test of the present formulation.

In this section we use the SkM$^*$ functional and optimize the configuration in a single Wigner--Seitz cell at fixed baryon density $n_\text{B}$ and proton fraction $Y_p$.
To reduce the computational cost, we neglect the spin-orbit term and take spin-degenerate single-particle orbitals as the output of the NN.
Whereas for finite nuclei the density is guided into a localized shape by pretraining, for the pasta phases the shape is guided by adding to the loss function a constraint term $\mathcal{L}_\text{ext}$ implementing a guiding potential appropriate to each structure, following Ref.~\cite{schuetrumpf2019}.
The explicit forms used here are given in Appendix~\ref{sec:otherformula}.
Because the shape is imposed in this way, the solutions reported below are the optimal configurations \emph{for each given shape}, not necessarily the global ground state at that density; determining which shape is energetically favored at a given $n_\text{B}$ requires comparing the converged energies of all candidate structures, which we leave to future work.

We compute three representative pasta phases, namely spheres, rods, and slabs, with the settings listed in Table~\ref{tab:pasta}.
The rod phase is generally expected around $n_\text{B}=0.045$--$0.055$~fm$^{-3}$ and the slab phase around $n_\text{B}=0.07$~fm$^{-3}$, but a higher baryon density requires a larger particle number and hence a greater computational cost.
We therefore restrict the present calculations to a lower-density region, where each shape is stabilized by the guiding potential; the purpose here is to demonstrate the representational capability of the ansatz rather than to locate the phase boundaries.
For the same reason we adopt a proton fraction of $Y_p=0.30$, which is considerably larger than the values of $Y_p\sim0.02$--$0.05$ expected in the inner crust, in order to keep the number of neutron orbitals tractable.
The actual particle numbers are set to the integers nearest to the real-valued particle numbers obtained from $n_\text{B}$ and $Y_p$.

\begin{figure*}[t]
    \centering
    \includegraphics[width=0.31\textwidth]{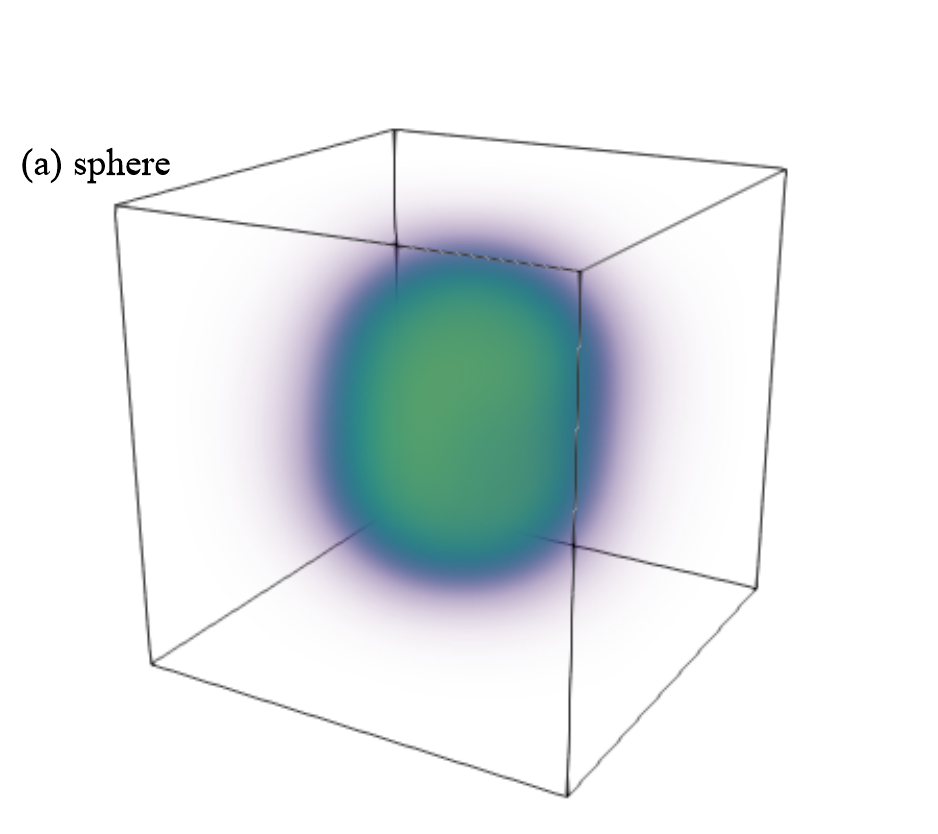}\hfill
    \includegraphics[width=0.31\textwidth]{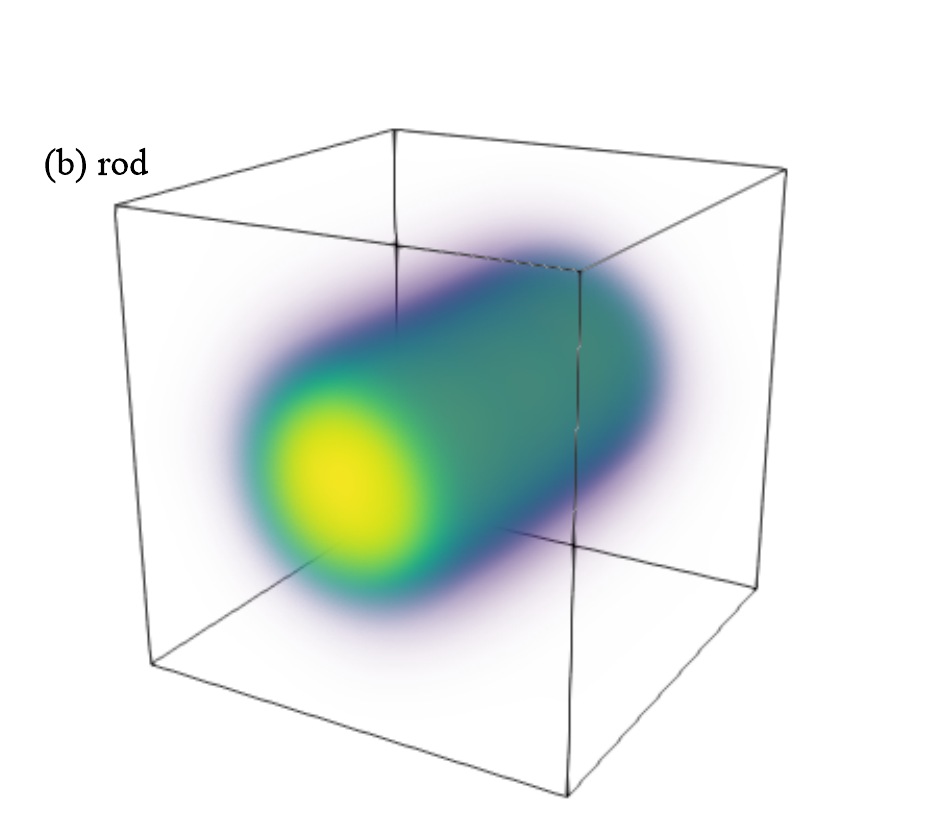}\hfill
    \includegraphics[width=0.31\textwidth]{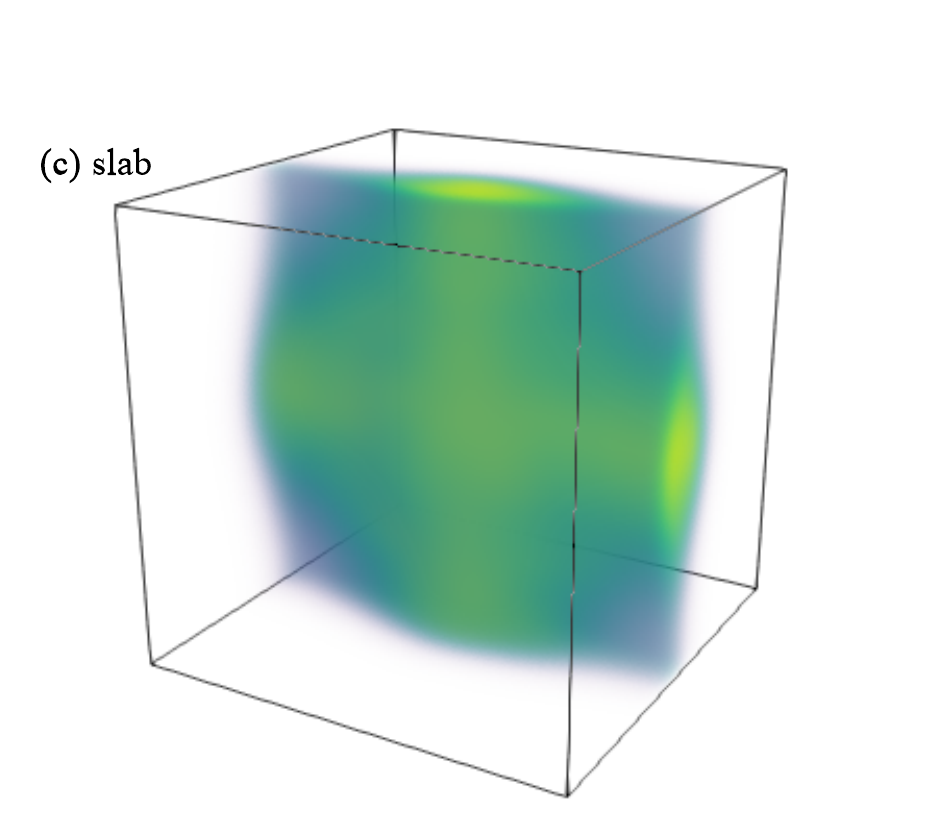}
    \caption{Nuclear density distributions of the pasta phases (a) sphere, (b) rod, and
    (c) slab, obtained with the neural-network method and the SkM$^*$ functional.}
    \label{fig:pasta_density}
\end{figure*}

The density distributions of the resulting configurations are displayed in Fig.~\ref{fig:pasta_density}. Figures~\ref{fig:pasta_density}(a), \ref{fig:pasta_density}(b), and \ref{fig:pasta_density}(c) correspond to the sphere, rod, slab phases, respectively, obtained with the parameters listed in Table~\ref{tab:pasta}. From the figure, we find that the single-particle wave functions based on the NN ansatz reproduce the density distributions of all three phases.
As discussed in Sec.~\ref{sec:nn_orbitals}, this reflects the adaptive nature of the MLP parametrization: during the energy optimization the basis functions themselves deform, so that the same architecture can represent density profiles ranging from localized finite nuclei to extended periodic structures.

\begin{table}[t]
    \centering
    \caption{Computational settings for each pasta configuration. From left to right: the
    configuration type, baryon density, proton fraction, size of the Wigner--Seitz cell, and
    proton and neutron numbers.}
    \label{tab:pasta}
    \small
    \setlength{\tabcolsep}{4pt}
    \begin{tabular}{lcccc}
        \toprule
        Phase & $\rho_B$ [fm$^{-3}$] & $Y_p$ & $L_{\text{cell}}$ [fm] & $(Z,N)$ \\
        \midrule
        Sphere & 0.020 & 0.30 & 14 & $(16,38)$ \\
        Rod    & 0.025 & 0.30 & 14 & $(20,48)$ \\
        Slab   & 0.030 & 0.30 & 18 & $(52,122)$ \\
        \bottomrule
    \end{tabular}
\end{table}

\section{Summary and Outlook}
\label{sec:summary}

In this work we extended the neural-network-based variational method proposed in our previous study~\cite{yoshimura2026nnetf} to the Kohn--Sham scheme, developing a framework in which the single-particle orbitals themselves are taken as the output of the network.
We first formulated the variational problem under the NN ansatz.
We showed that the variational condition of density functional theory (DFT) is realized as a projection onto the tangent space of the manifold spanned by the NN parameters, and that the energy minimization performed in deep learning corresponds to the simultaneous optimization of the basis functions and of the coefficients of the expansion in that basis.
We then presented the Skyrme energy density functional (EDF) including the spin-orbit terms, and assessed the method through three-dimensional calculations.

The method was validated in three steps.
First, for the closed-shell nucleus $^{16}$O, we varied the number of units, the number of layers, and the arithmetic precision, and examined systematically how the results and the cost per step respond.
We found that a moderate depth combined with a moderate width is required for quantitative accuracy, and that single precision suffices to represent the nuclear density distribution---a property that should become advantageous on forthcoming AI-oriented hardware, where low- and mixed-precision arithmetic is the natural operating point.
Second, for closed-shell nuclei from $^{16}$O to $^{208}$Pb, we computed the total energy and the charge radius and found agreement with conventional Skyrme--Hartree--Fock results at the level of $0.3\%$ for the charge radii and $0.8\%$ for the energies per nucleon.
We also computed open-shell deformed nuclei such as $^{20}$Ne, $^{24}$Mg, and $^{238}$U, and obtained deformed ground states in reasonable agreement with reference calculations, although the present framework does not yet include pairing contributions.
Third, applying the method to the nuclear pasta phases, we demonstrated that single-particle orbitals generated by the NN ansatz can represent all three fundamental shapes---spheres, rods, and slabs---within a Wigner--Seitz cell.

The present method is significant in two respects.
First, for extensions of the EDF that are actively studied at present, such as tensor terms and higher-order gradient terms, the ground state can be obtained by optimizing the functional directly, without analytically deriving the corresponding single-particle Hamiltonian.
Although we have not exploited this feature in the present work, it should reduce considerably the effort required to explore new functional forms.
Second, the formulation inherits developments in machine-learning infrastructure without any special adaptation, including advances in GPU hardware, automatic differentiation, and mixed-precision arithmetic.

A natural next step is the extension to DFT including pairing, that is, to the Hartree--Fock--Bogoliubov (HFB) theory or superfluid local density approximation (SLDA) in Kohn-Sham DFT.
This requires representing the quasiparticle amplitudes $U$ and $V$ rather than a single set of single-particle orbitals, and retaining a sufficiently large quasiparticle space including continuum states.
The principal technical difficulty is to impose the quasiparticle orthonormality conditions on the network output, which are more involved than the orthonormality constraint treated here; how best to enforce them within the NN ansatz remains to be established.
If a GPU-oriented formulation can be developed for HFB theory, it would open a new avenue for studies of nuclear structure including pairing correlations, complementing the iterative solvers that have been refined for this purpose~\cite{jin2020cocg, kashiwaba2020cocr}.
Extensions to more complex systems, such as multi-configurational Hartree--Fock theory involving many Slater determinants~\cite{matsumoto2023mchf, matsumoto2025mchf}, are also possible.
More broadly, combining NN representations of the single-particle wave functions with automatic differentiation may open the way to applications in time-dependent Hartree--Fock theory and the random-phase approximation that allow for efficient calculations of nuclear excitations, reactions and dynamics.
The framework established here provides a foundation for such extensions and, more generally, for nuclear-theory calculations on the computing platforms of the coming decade.

\section*{Acknowledgement}
This work is financially supported by the JSPS Research Fellow, Grant No.~JP24KJ1110 (K.Y.).
This work is also supported by JSPS Grant-in-Aid for Scientific Research KAKENHI Grants
No. JP23K03410, No. JP23K25864, and No. JP25H01269 (K.S.).
This work used the computational resources of TSUBAME4.0 at Center for Information Infrastructure in Institute of Science Tokyo (Project ID: hp250097), and Miyabi at Joint Center for Advanced High Performance Computing (JCAHPC) in the University of Tokyo (Project ID: hp260222), through the HPCI System Projects.

\appendix 

\section{Other Formulae and Observables}
\label{sec:otherformula}

\subsection{Orthonormalization with Cholesky Factorization}
We denote the wave function computed through the MLP and multiplied by an appropriate envelope as
\begin{equation}
    \tilde{\psi}_{q,i}(\bm{r}) = \text{MLP}_{q,i}(\bm{r}) g_\text{env}(\abs{\bm{r}}) .
\end{equation}
This wave function is in general not orthonormal.
One efficient way to orthonormalize such a set of vectors is to use the Cholesky factorization.
We first compute the overlap matrix
\begin{equation}
    S_{ij} = \braket{\tilde{\psi}_i}{\tilde{\psi}_j} = \sum_\sigma \int d\bm{r}\, \tilde{\psi}_i^*(\bm{r}\sigma) \tilde{\psi}_j(\bm{r}\sigma) .
\end{equation}
Decomposing this matrix with a lower-triangular matrix $L$ as
\begin{equation}
    S = LL^\dagger ,
\end{equation}
the new set of orbitals defined by
\begin{equation}
    \psi_i = \sum_j \qty(L^{-1})_{ij} \tilde{\psi}_j
\end{equation}
satisfies the orthonormality condition.
As shown in the main text, the computational framework of this work operates without problems even with FP32 variables.
However, computing the overlap matrix for this Cholesky factorization in single precision causes rounding errors to accumulate, making the matrix non-positive-definite and causing the factorization to break down.
Therefore, even in the FP32 calculations of this work, we compute only the overlap matrix and the factorization in FP64.

\subsection{Charge Radii}
The charge radius is evaluated using the formula~\cite{bertozzi1972charge} that is also employed in the protocol of the SLy series~\cite{chabanat1998sly}.
Its general expression is written as
\begin{equation}
    \expval{r^2}_\text{ch} = \expval{R^2}_p^\text{pt} + \expval{r^2_p} + \frac{N}{Z}\expval{r_n^2} + \expval{r^2}_\text{SO} .
\end{equation}
The first term is the mean-square radius for point-like protons,
\begin{equation}
    \expval{R^2}_p^\text{pt} = \frac{1}{Z}\int \dd\bm{r}\,\rho_p(\bm{r}')\abs{\bm{r}'}^2 ,
\end{equation}
where $\bm{r}'$ is the coordinate measured from the center of mass,
\begin{equation}
    \bm{r}' = \bm{r} - \bm{R}_\text{cm},\quad \bm{R}_\text{cm} = \frac{1}{Z}\int \dd\bm{r}\, \rho_p(\bm{r})\bm{r} .
\end{equation}
The second and third terms are the single-nucleon charge contributions of the proton and neutron, for which we use
\begin{equation}
    \expval{r^2_p} = 0.64\,\fm^{2},\quad \expval{r^2_n} = -0.1161\,\fm^{2} .
\end{equation}
The fourth term is the spin-orbit contribution,
\begin{equation}
    \expval{r^2}_\text{SO} = -\frac{1}{Z} \sum_q \frac{\mu_q}{m_N^2} \int \dd\bm{r}\, \bm{r}'\cdot \bm{J}_q(\bm{r}) ,
\end{equation}
where $\mu_q$ is the nucleon magnetic moment, for which we use
\begin{equation}
    \mu_p = +2.793,\quad \mu_n = -1.913 .
\end{equation}

\subsection{Deformations}
As a general formulation for describing nuclear deformation, one commonly uses the multipole moments~\cite{bizzeti2004deform, agbemava2016deform}
\begin{equation}
    Q_{\ell m} = \int \dd\bm{r}\, \hat{Q}_{\ell m}\rho_0(\bm{r}),
\end{equation}
where $\ell=2$ corresponds to the quadrupole and $\ell=3$ to the octupole deformation.
In typical Skyrme Hartree--Fock calculations, one usually assumes a deformation about the $z$ axis and evaluates the quadrupole deformation using
\begin{equation}
    \hat{Q}_{20} = 2z'^2 - x'^2 - y'^2 ,
\end{equation}
where $x'$, $y'$, and $z'$ are, as in the previous subsection, the coordinates relative to the center of mass.
In the neural-network-based method, on the other hand, the energy is minimized in fully three-dimensional space without assuming any axis, so that in general the direction of the deformation axis is not known a priori.
Indeed, Fig.~\ref{fig:deformed_density} shows that the deformation axis does not coincide, at least, with the $z$ axis.
We therefore first compute the general quadrupole tensor
\begin{equation}
    Q_{ij} = \int \dd\bm{r}\, \rho_0(\bm{r})\qty(3 x'_i x'_j - \abs{\bm{r}'}^2 \delta_{ij}),
\end{equation}
where $x_i\in \{x, y, z\}$.
Since this is a real symmetric $3\times 3$ matrix, there exists an orthogonal matrix $O$ that diagonalizes it,
\begin{equation}
    O^\mathtt{T} Q O = \mqty(\dmat[0]{q_1, q_2, q_3}).
\end{equation}
The resulting eigenvalues $q_i$ represent the deformation about the principal axes.
In this work, we identify the largest eigenvalue with the quadrupole deformation $Q_{20}$.
For the actual deformation parameter $\beta_2$, we use the dimensionless quantity
\begin{equation}
    \beta_2 = \frac{\sqrt{5\pi}}{3}\frac{Q_{20}}{A R_0^2},\quad R_0 = 1.2A^{1/3}\,\fm .
\end{equation}
We denote the deformation axis obtained here by $\bm{n}$.
For an axially symmetric shape, the octupole and quadrupole deformations share the symmetry axis, so the octupole deformation can be examined by looking at the deformation about the axis $\bm{n}$.
Projecting the coordinate onto this axis as $z_n = \bm{r}'\cdot\bm{n}$, we obtain
\begin{equation}
    Q_{30} = \int \dd\bm{r}\, \rho_0(\bm{r})\qty(5z_n^3 - 3z_n\abs{\bm{r}'}^2) ,
\end{equation}
and the normalization to the deformation parameter $\beta_3$ is
\begin{equation}
    \beta_3 = \frac{\sqrt{7\pi}}{3AR_0^3}Q_{30} .
\end{equation}
The calculation results of these deformation parameters are discussed in Sec.~\ref{sec:finite_nuclei}.

\section{Canonical Kohn--Sham orbitals}
\label{sec:canonical_orbitals}

\begin{table}[tp]
\centering
\caption{Single-particle energies $\varepsilon_{n\ell j}$ (in MeV) of the occupied
neutron (n) and proton (p) levels of $^{16}$O and $^{40}$Ca obtained from the
NN--Skyrme (SLy4) solution. The levels are the eigenstates of the mean-field
Hamiltonian $\hat h=\delta E/\delta\psi^{*}$ diagonalized in the occupied subspace.}
\label{tab:spe}
\begin{tabular*}{0.7\columnwidth}{@{\extracolsep{\fill}}cccc}
\toprule
Nucleus & Orbital & $\varepsilon_n$ & $\varepsilon_p$ \\
        &         & (MeV) & (MeV) \\
\midrule
$^{16}$O
        & $1s_{1/2}$ & $-36.15$ & $-32.36$ \\
        & $1p_{3/2}$ & $-20.56$ & $-17.08$ \\
        & $1p_{1/2}$ & $-14.54$ & $-11.17$ \\
\midrule
$^{40}$Ca
        & $1s_{1/2}$ & $-48.30$ & $-40.29$ \\
        & $1p_{3/2}$ & $-35.01$ & $-27.47$ \\
        & $1p_{1/2}$ & $-31.10$ & $-23.65$ \\
        & $1d_{5/2}$ & $-22.01$ & $-14.86$ \\
        & $2s_{1/2}$ & $-17.26$ & $-10.15$ \\
        & $1d_{3/2}$ & $-15.31$ & $-8.34$ \\
\bottomrule
\end{tabular*}
\end{table}

\begin{figure}[tp]
    \centering
    \includegraphics[width=0.4\textwidth]{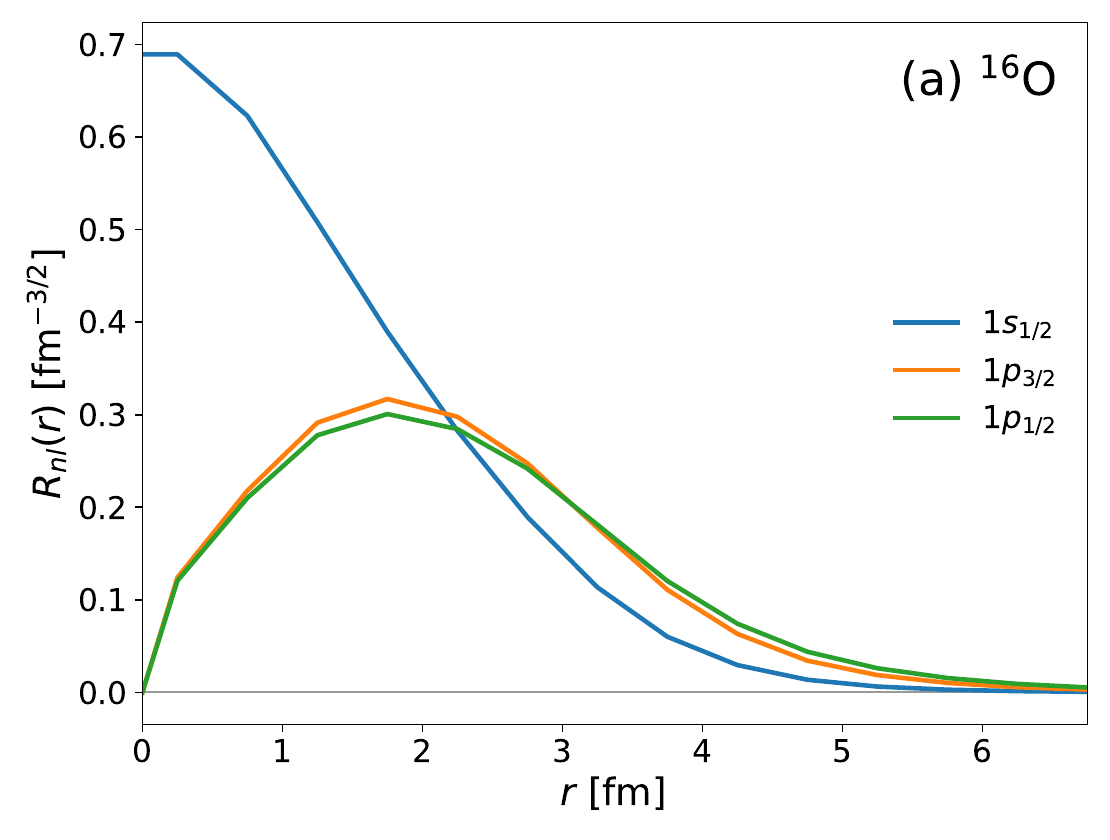}\\
    \includegraphics[width=0.4\textwidth]{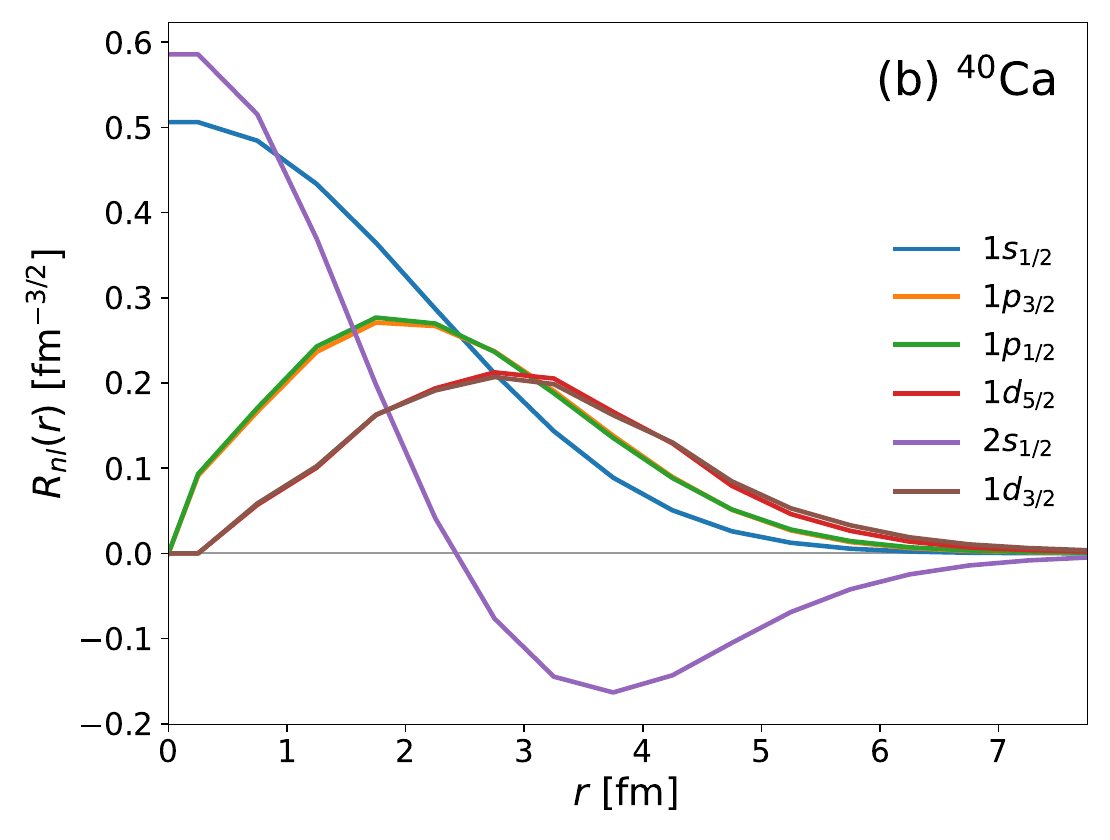}
    \caption{Radial distribution of neutron single-particle orbitals of (a)$^{16}$O and (b)$^{40}$Ca calculated by diagonalizing the mean-field Hamiltonian with the neural-network orbitals.}
    \label{fig:eigenstate}
\end{figure}

The neural-network orbitals after the orthonormalization procedure constitute, as a whole, the nuclear density distribution; however, they do not directly correspond to single-particle orbitals, that is, to eigenfunctions of the single-particle Hamiltonian.
Nevertheless, once the set of NN orbitals in the ground state has been obtained, we can determine the energy eigenvalues and eigenfunctions through a suitable procedure.
We construct the Hamiltonian kernel between the levels as
\begin{equation}
    H_{ij}^{(q)}
    =
    \mel{\psi_{q,i}}{\hat h_q}{\psi_{q,j}} = \int \dd\bm{r}\, \psi^*_{q,i}(\bm{r}) \hat{h}_q(\bm{r})\psi_{q,j}(\bm{r}),
    \label{eq:occupied_hamiltonian_matrix}
\end{equation}
where $\hat{h}_q$ denotes the single-particle Hamiltonian.
Although in this work the Hamiltonian is not formulated explicitly, from the variational condition, one has
\begin{equation}
    \hat h_q\psi_{q,j}(\rvec)
    =
    \frac{\delta E}{\delta\psi_{q,j}^*(\rvec)}
    \label{eq:autograd_hamiltonian}
\end{equation}
which can be easily evaluated using the automatic-differentiation functionality provided in standard machine-learning packages such as PyTorch.
By diagonalizing the Hamiltonian kernel $H^{(q)}_{ij}$ constructed in this way, we can obtain the set of eigenvalues and eigenfunctions.
The size of this matrix equals the number of orbitals, namely the particle number (or half of it if spin degeneracy is assumed), which is at most a few hundred, so the diagonalization is easy to perform.
The energy levels computed for $^{16}$O and $^{40}$Ca, together with plots of the neutron single-particle orbitals, are shown in Table~\ref{tab:spe} and Figure~\ref{fig:eigenstate}.
As can be seen, both the energy eigenvalues and the eigenstates of each nucleus fall within physically reasonable solutions.

\clearpage

\bibliographystyle{apsrev4-2}
\bibliography{shfml}

\end{document}